\documentclass{sf2a-conf2025}
\usepackage{graphicx}
\usepackage{hyperref}
\usepackage[]{natbib}  
\usepackage{epstopdf}

\def\BibTeX{{\rm B\kern-.05em{\sc i\kern-.025em b}\kern-.08em
    T\kern-.1667em\lower.7ex\hbox{E}\kern-.125emX}}
\bibpunct{(}{)}{;}{a}{}{,}  

\begin{document}

\TitreGlobal{SF2A 2025}


\title{A few ideas to promote inclusion}

\runningtitle{Ideas for inclusion}

\author{T. Paumard}\address{LIRA, Observatoire de Paris, Université PSL, CNRS, Sorbonne Université, Université Paris Cité, 5 place Jules Janssen, 92195 Meudon, France}

\author{A. Guilbert-Lepoutre}\address{LGL-TPE, CNRS, Universit´e Lyon 1, ENSL, Villeurbanne, France}

\author{M. Clavel}\address{Univ. Grenoble Alpes, CNRS, IPAG, 38000 Grenoble, France}

\author{F. Cornu$^1$}

\author{L. Petitdemange$^1$}

\author{F. Dulieu$^1$}

\author{L. Griton$^1$}

\author{R.-M. Ouazzani$^1$}

\setcounter{page}{237}


\maketitle


\begin{abstract}
Promoting diversity, equity, inclusion and accessibility (DEIA) is
both a legal and professional responsibility in French research
institutions. This paper presents practical strategies to foster
inclusive work environments within French research units. We summarize
the regulatory context, key findings from the INSU-AA prospective on
discrimination, and fundamental principles for promoting equity. We
discuss approaches to mitigate implicit biases across all career
stages, from early education to retirement, and outline strategies for
equitable recruitment and career advancement. Concrete initiatives in
one of our units (LESIA/LIRA) are described, including internal
communications, exhibitions, and accessible pedagogical
activities. The creation of a dedicated commission within the unit
council ensures coordinated DEIA efforts, legitimized by institutional
support and methodical planning. By sharing these experiences, we
provide actionable guidance for research units seeking to advance DEIA
in science.
\end{abstract}



\section{Introduction}

\emph{Liberté, Égalité, Fraternité}: such is the Motto of the French
Republic. This motto resonates with the ever-growing desire within
civil society, in France and around the world, to see these principles
flourish in general as well as in the context of the world of work.

Several works have demonstrated the benefit of nurturing certain forms
of liberty at work: for instance \citet{Spector1986} found in a
meta-analysis that high levels of perceived control at work (i.e. job
autonomy) were linked to greater job satisfaction, commitment,
motivation, and lower emotional distress, absenteeism, turnover intent
and turnover, and physical symptoms. It goes well beyond the scope of
these proceedings to discuss further how the many aspects of the
concept of Liberty are relevant at work, though.

The last word of the motto, \emph{Fraternité}, conveys the idea that
people should behave with each other like siblings in a healthy
familly. But some authors have raised questions concerning its usage
in the motto, on the basis that it may be understood as excluding
\emph{sisters}: in an advisory \citep{HCE2018}, the French \emph{Haut
conseil à l'Égalité entre les femmes et les hommes} (HCE, High council
for Equality between women and men) estimates, with Réjane Sénac
\citep{Senac2017-he} that ``in the motto `Liberty, Equality,
Fraternity,' the term \emph{fraternity} expresses not republican
neutrality, but the historical and legal exclusion of women from the
political community''\footnote{The motto is part of the preamble of
the French consistution since 1848, almost one century before women
obtained the right to vote in 1944.}. For this reason, the HCE
recommends to initiate a reflection about the usage of this word in
the French constitution where it could be replaced by a gender-neutral
alternative: either \emph{solidarité} (solidarity) which is common but
is not a strict gender-neutral equivalent as it lacks the notion of
familly, or the more sophisticated \emph{adelphité} (siblinghood).

The HCE also argues that ``[{solidarity}] has the advantage of no
longer placing the political community in relation to a familial or
religious community'', again raising the question of the boundaries of
such a community. Regardless of their opinion about the wording of the
French motto, the individuals or research units who care for
\emph{Égalité} (Equality) in their environment are immediately faced
with this question. One possible answer is given by referring to the
Law: the French Penal Code and Labor Code define at least 25 unlawful
grounds of discrimination including, to name a few, origin, sex,
disability, sexual orientation, gender identity, religion and
ethnicity\footnote{\url{https://www.service-public.fr/particuliers/vosdroits/F38175};
Penal Code 225-1 to 225-1-2, Labor Code L1134-1}.

In this paper, we will share ideas from which an individual or group
of people (especially a French research unit) can take inspiration in
order to foster diversity, equity, inclusion and accessibility (DEIA) in
their work environment, trying to keep these ideas applicable to any
of the unlawful grounds of discrimination according to French Law. We
will first remind a few points of context in
Sect.~\ref{Paumard:context}, including the French regulatory
framework, a few results from the INSU prospective and basic
principles to foster a inclusive work environment. In
Sect.~\ref{Paumard:fostering-diversity}, we will discuss how implicit
biases affect women and marginalized groups from preschool to
retirement, and share ideas to counter them. We will then focus in
Sect.~\ref{Paumard:lesia-lira} on the actions undertaken in one of our
units, in the hope that they may be useful to others.

\section{Context}
\label{Paumard:context}

\subsection{A regulatory framework for institutional action}

\label{Paumard:law}

The \emph{Mission pour la Place des Femmes (MPDF)} of the CNRS was
created in 2001 in order to promote gender equality and strengthen the
position of women within French research. It was one of the first
equality bodies in a public research institution in France. Initially
focused on identifying gender imbalances and raising awareness, the
Mission gradually expanded its role to include the production of
statistical data, the implementation of action plans, and the support
of equality initiatives at national and European levels. Today, it
plays a central role in driving institutional policy on gender
equality inside the CNRS, coordinating training, monitoring progress,
and working to combat discrimination and stereotypes in the research
community.

A regulatory framework applies to French public employers with regard
to discrimination, particularly concerning gender and disability.

French public employers are subject to a series of legal obligations
regarding professional equality between women and men. Since the Law
of 13 July 1983, they are required to guarantee equal treatment in
access to employment, training and career development. The Law of 9
May 2001 introduced the obligation to include gender equality
objectives in annual staff reports and in social dialogue. The Law of
6 August 2019 established the obligation for public employers to adopt
and implement a gender equality action plan, based on published
indicators and concrete corrective measures.

French public employers are also subject to legal obligations
concerning the employment of persons with disabilities. The Law of 10
July 1987 introduced a mandatory employment quota of 6\% of workers
with disabilities in all public and private organisations. The Law of
11 February 2005 strengthened this obligation, broadened the
definition of disability and required public employers to implement
measures to facilitate recruitment, accessibility, retention and
workplace adaptation. The Law of 6 August 2019 further reinforced
these obligations.

In this context, the Astronomy and Astrophysics domain of INSU
(\emph{Institut national des sciences de l'Univers}) has created in
2021 a network of Equality correspondents with the aim of gathering
two persons in each laboratry of the domain \citep{Clavel2024}. In a
nutshell, their missions are:
\begin{itemize}
\item Listening to and guiding victims and witnesses of sexual and
  gender-based violence (SGBV) (the correspondents are trained for
  this), moral or sexual harassment, and discrimination on any illegal
  grounds;
\item Advising management and communicating within the unit to promote
  gender parity and professional equality, prevent sexual violence,
  harassment, and discrimination, and raise awareness of these issues
  among all employees.
\end{itemize}

\subsection{Conclusions of the INSU prospective}

As part of the INSU 2024 Astronomy and Astrophysics (AA) prospective
exercise, Working Group (WG) I.1 was tasked with assessing the state
of inclusion, diversity, and equality within the community, and with
recommending improvements where necessary
\citep{GuilbertLepoutre2024}.

The French AA community was surveyed on a broad range of self-reported
experiences of inequality and discrimination, including gender-based
discrimination, racism, homophobia, ableism, and social status. While
the survey results have limited statistical weight, they are fully
consistent with larger-scale studies—both in the general population
(for specific forms of discrimination) and within academia (across
various disciplines and at the international level).

Key findings are summarized below:
\begin{itemize}
    \item Only $42\%$ of respondents reported never having been
      targeted by any form of discrimination.
    \item Each type of discrimination was personally experienced by
      approximately $10$--$20\%$ of respondents overall.
    \item However, these figures increase sharply among individuals
      belonging to marginalized groups. WG I.1 was able to demonstrate
      statistical differences between targeted groups and the rest of
      the community. For instance:
      \begin{itemize}
      \item Only $\simeq 20\%$ of women reported never having
        experienced sexist behavior.
      \item Only $\simeq 25\%$ of LGBTQ+ respondents reported never
        having experienced homophobia.
      \item Around $50\%$ of disabled respondents reported never
        having been targeted by ableist behavior.
      \end{itemize}
\end{itemize}

The survey also investigated the prevalence of SGBV in the
workplace. Based on self-reports, WG I.1 estimates that $\simeq 20\%$
of women have experienced SGBV at least once in their workplace,
including reports of aggressions and rape. Similarly, at least $20\%$
of LGBTQ+ respondents reported experiencing SGBV, in line with the
\emph{Baromètre} \citet{IFOP2024}, which found that three out of ten LGBTQ+
employees in the general population had suffered at least one form of
workplace aggression. Even more concerning, the most recent cases of
SGBV ($<5$ years, including at least one rape) exclusively involved
young, non-tenured female colleagues.

Overall, these results underscore the urgent need to sustain efforts
in favor of gender equity and the fight against SGBV, while extending
such measures to other forms of discrimination.

\subsection{A few basic principles}

Here we remind basic principles that apply in general for the
promotion of inclusion in an academic unit:

\begin{itemize}
\item Make it a topic. Use every opportunity to raise awareness
  (e.g. general assemblies, research unit councils etc.).
\item Promoting inclusion is a long-distance race, not a sprint. To
  avoid backlash that could jeopadize your efforts on the long run, be
  mindful of the acceptability of your action. Ensure its legitimacy
  by basing it on solid ground, such as the legal and institutional
  framework exposed above.
\item Get trained. The supervisory bodies of research units (including
  CNRS) all offer training programs on a variety of subjects,
  including inclusion. In addition, the Ministry of Higher Education
  and Research (ESR) in collaboration with
  Coop-Egal\footnote{\url{https://www.youtube.com/@CoopEgal385}}
  offers a series of webinars on equality every year (EGA-ESR).
\item Respond to instances of discriminatory behavior or verbal abuse
  in meetings that you may witness. To this effect, you can apply
  Right to Be's \emph{5Ds of bystander intervention} \citep{FiveDs}:
\begin{description}
  \item[Distract:] create a diversion to allow for deescalation,
    e.g. pretend you have misplaced your keys;
  \item[Delegate:] if you feel something needs to be done but you are
    not comfortable stepping in yourself, ask someone else to
    intervene who might be better positioned to do so;
  \item[Document:] if someone else is already helping, take notes or
    record (always ask the harrased person what \emph{they} want to do
    with this documentation);
  \item[Delay:] if you could not act immediately, providing help and
    support after the fact can make a difference;
  \item[Direct:] directly intervene, e.g. by asking the victim
    if they are okay, or by addressing the perpetrator.
\end{description}
\item Do not assume that you are unbiased. Everyone has
  biases\footnote{You may explore your implicit biases at:
  \url{https://implicit.harvard.edu/}.}. While it is not possible to
  eliminate them entirely, acknowledging your own biases fosters the
  self-awareness needed to notice, pause, and counteract their
  influence \citep[e.g.][]{Devine2012}.
\item Prepare, implement and follow an action plan on DEIA matters (see
  Sect.~\ref{Paumard:action-plan}).
\item Create a DEIA team. A formal ``DEIA'' or ``Professional Equality''
  team, commissioned by the unit council, will help in several aspects
  mentionned above. Its action is de facto legitimated, and it will
  automatically ``make it a topic'' since the commission will report
  regularly in front of the council. This commission can also
  structure a professional training plan around DEIA topics as well as
  an action plan.
\end{itemize}

Such efforts align with national recommendations: for instance, the
\emph{Kit de prévention des discriminations dans l’enseignement
supérieur} developed by the \emph{Défenseur des droits} and
institutional partners provides concrete measures to foster equity in
a research environment \citep{defenseur2021kit}.

\section{Fostering diversity from preschool to retirement}

\label{Paumard:fostering-diversity}

The metaphor of the leaky pipe illustrates that women leave academic
research (or that their career ceases to progress) at every stage of
their career, with $57\%$ of general baccalaureate graduates, $38\%$
of research assistants and $29\%$ of CNRS research directors being
females \citep{MPDF2021}. In addition, research has shown that
gendered stereotypes emerge as early as the preschool years and
discourage young girls from pursuing future scientific or engineering
careers \citep{Bian2017,Master2021}. Similarly, studies have
demonstrated that ethnicity-based biases are acquired at very young
stages \citep{Meltzoff2024}. In the absence of comparable statistics
for other grounds of discrimination, it is legitimate to postulate
that similar trends also apply to them and this calls for actions
targetting every stage of life, from school and even preschool years
to retirement.

Many researchers in our field engage regularly in science
dissemination activities towards children and teenagers from preschool
to high school. Ideally, it would be wonderfull to see a very diverse
population of members of our community do so, but as long as our
community is \emph{not} diverse, we need to do specific efforts to
pass the message that science is for everyone, independant of gender,
ethnicity, social origin or anything else. Such efforts can be
implemented on an individual or collective basis:
\textbullet\ make every effort possible to cite women scientists;
\textbullet\ include a slide showing the diversity of your team at least in
  terms of gender and ethnicity;
\textbullet\ use this slide to introduce the various statuses of our
  colleagues (researchers but also engineers, technicians and
  administrative staff).

Another way to foster equity and diversity in our science
communication practice is to engage with underserved
audiences---including people with disabilities---and marginalized
communities, such as those in priority education zones and rural
areas. Platforms like
ViensVoirMonTaf\footnote{\url{https://www.viensvoirmontaf.fr/}} can
help proposing internships to pupils in the tenth and eleventh grades
(\emph{classes de troisième et de seconde} in the French system) in
these marginalized areas, while events such as Cercle FSER's
\emph{declics}\footnote{\url{https://www.cerclefser.org/fr/declics/}}
can help organizing visits in schools, with the possibility to
prioritize on communities with special needs. Specific exhibitions can
be designed to be accessible to people with certain disabilities (see
an example in Sect.~\ref{Paumard:blind}).

Universities are generally well aware of diversity and inclusion
matters. Points of attention include implicit biases and care for
diversity during selection processes, display of diversity on the
website and other material on which prospective students may base
their choice, and visible anti-harassment policies. The ``Aspie
Friendly''
program\footnote{\url{https://handicap.gouv.fr/aspie-friendly}} fosters
inclusion of autistic students.

When recruiting people in our units (whatever the status), we should
consider the following points:
\textbullet\ publicly communicate an inclusion policy (it enhances
  attractiveness);
\textbullet\ adopt a code of conduct and make it publicly available;
\textbullet\ write genuinely stereotype-free job postings (note: this
  requires training);
\textbullet\ avoid indirect discrimination---e.g., do not require attendance
  at elite institutions;
\textbullet\ educate oneself and one's team about unconscious biases;
\textbullet\ use explicit, needs-based, non-discriminatory evaluation
  criteria;
\textbullet\ be mindful of the recommendation letters we write---and the
  weight we give to we read---as letters of recommendation
  written for women, even when written in good faith with the hope of
  helping the applicant, are often tinged with unconscious biases that
  work against them \citep[e.g.][]{Chang2023};
\textbullet\ ensure information reaches everyone---beware of the ``old boys'
  club'' effect by which crucial information such as job opportunities
  sometimes is shared only during informal events (afterworks, coffee
  breaks...), never reaching those who do not participate to such
  events.

Once recruited, colleagues must be provided with equal opportunities
for career development, through fair evaluation and promotion:
\textbullet\ encourage colleagues to apply for promotions, prizes, fellowships,
  and other distinctions, particularly those who may experience
  impostor syndrome, which disproportionately affects women and
  underrepresented groups \citep{
    clance1978imposter};
\textbullet\ consider specific circumstances (e.g., disability, maternity,
  caregiving responsibilities) in the evaluation of colleagues;
\textbullet\ ensure promotion decisions reflect the available talent pool,
  rather than only the applicant pool; research shows that men are
  often more likely to apply even with less advanced application files
  \citep[e.g.][]{ABRAHAM2024348}.

Note that some promotion pathways for engineers, technicians, and
administrative staff go through the research unit's council. Raising
awareness within the council about fair evaluation is therefore very
important. This can be an action item for a DEIA team.

\section{Examples of actions led at LESIA/LIRA}

\label{Paumard:lesia-lira}

Now that we have established some general principles, we turn to
concrete examples of actions implemented since the creation of the
INSU-AA Equality Network in late 2021, through the correspondent
appointed by the LESIA research unit at Paris Observatory–PSL. On
January 1$^\mathrm{st}$, 2025, LESIA merged with parts of GEPI and LERMA
to form a new research unit, the \emph{Laboratoire d'instrumentation
et de recherche en astrophysique} (LIRA). The seeds planted by the
Equality correspondents in the three former laboratories continue to
flourish at LIRA, as illustrated below.

One of the first initiatives was to update the unit’s intranet with
resources for witnesses and vitims of SGBV as well as moral or sexual
harassment. This section of the intranet also introduces the network
correspondents and clarifies their role. In parallel, we sought to
provide colleagues with accessible information on equality in a
professional context. LESIA, and now LIRA, circulates a weekly
internal newsletter to all members, in which we regularly include
(every one to three months) a short note on Equality. These notes
typically consist of a brief introduction and a pointer to an external
online resource, such as a video, leaflet, or announcement of an
upcoming event. All previous notes are archived on the intranet,
alongside the more institutional information mentioned above.

We then initiated a small collection of relevant books, primarily in
French but also including several works in English, which we make
available to colleagues and students through an \emph{Equality
Library}. To publicize this library and facilitate connection with
unit members, we organize informal \emph{Equality Coffee Breaks} near
coffee machines, complete with croissants. These gatherings have
proven invaluable---not only for addressing reservations colleagues
may have about our initiatives, but also for making us more
approachable to quieter collaborators.

Another way we seek to engage colleagues and stimulate reflection and
discussion is through exhibitions or poster campaigns. We have
organized two so far. The first was inspired by the Hypatia
project\footnote{\url{https://www.femmesetsciences.fr/40-soeurs-hypatie}}:
while the Eiffel Tower bears the names of 72 French male scientists
engraved around the perimeter of its first floor, the project aims to
have the names of 40 female scientists inscribed on its second floor.
In the same spirit, we selected 40 female scientists and prepared for
each a short portrait combining a photograph and a brief text,
together with a QR code linking to further information. These
portraits were then displayed in the corridors of the unit.

The second campaign, organized for Pride Month, consisted of five
posters. Each provided information about LGBT-phobia (including the
definition of LGBT, the prevalence of LGBT-phobic attacks in the
workplace in France, etc.), accompanied by the portrait of a
scientific figure—historical or contemporary—who identifies or
identified as LGBT.

Still aiming at “making it a topic,” we strive to be present as
much as possible at the unit’s general assemblies. During the most
recent one, we organized an activity using pedagogical material
designed to present the Solar System to visually impaired audiences:
3D-printed planets with slightly accentuated surface reliefs, enabling
participants to identify each planet solely by touch. This activity
illustrates both how astronomy can be made accessible to underserved
audiences, and how disability can be addressed and discussed among our
colleagues.\label{Paumard:blind}

LIRA's unit council recently appointed a dedicated commission called
\emph{Égalités et qualité de vie au travail}, broader in scope than
the Equality correspondents alone. This commission is tasked with
writing a dedicated action plan\label{Paumard:action-plan} on equality
and monitoring its implementation. The plan aligns with the policies
and objectives defined by our supervisory bodies---for example, their
gender equality action plan mandated by Law (Sect.~\ref{Paumard:law}),
as well as their \emph{Schéma directeur développement durable et
responsabilité sociétale et environnementale}, mandated by the
Ministry of ESR. This institutional framework provides our actions
with a high degree of legitimacy and acceptability, and ensures that
we can “make it a topic” regularly before the unit council.

Finally, we stress that we are astronomers: we may not have been
formally trained to promote Equality, but we have been trained to
design, implement, and revise plans---something we routinely do when
building instruments. This familiar methodology should strongly
support us in achieving our goals.

\section{Conclusion}

Fostering diversity, equity, and inclusion in research units requires
continuous effort, from raising awareness and addressing biases to
implementing structured action plans. Practical initiatives---ranging
from thoughtfully crafted science talks in schools to training of
committees involved in career progression---can promote inclusion
across all career stages. Institutional support, formal DEIA teams, and
methodical planning ensure legitimacy and sustainability. The
experience at LESIA/LIRA illustrates how research units can translate
legal and ethical commitments into concrete, everyday actions.

\begin{acknowledgements}
We warmly thank the organizers of SF2A 2025 for a superbly organized
event as well as Coop-Egal for their very instructive training
sessions on professional equality. AI-based tools (ChatGPT, DeepL)
were used to help sourcing information and translating.
\end{acknowledgements}

\bibliographystyle{aa}  
\bibliography{Paumard_S14} 

@ARTICLE{Spector1986,
   author = {{Spector}, P. E.},
   journal = {Human Relations},
   year = 1986,
   volume = 39,
   pages = {1005-1016}
}

@misc{HCE2018,
  title        = "Pour une Constitution garante de l’égalité femmes-hommes",
  author       = "{Bousquet}, D. and {Sénac}, R. and {Gayraud}, A. and {Guiraud}, C.",
  year         = 2018,
  month        = apr,
  howpublished = "avis n°2018-04-18-PAR-033 publié le 18 avril 2018, Haut conseil à l'Égalité entre les femmes et les hommes, 55, rue Saint-Dominique - 75007 Paris, \url{https://www.haut-conseil-egalite.gouv.fr/IMG/pdf/hce_rapport_constitution-garante-v4.pdf}"
}

@misc{GuilbertLepoutre2024,
  title        = "Rapport de synthèse des travaux du groupe I.1: Inclusion, Diversité, Egalité",
  author       = "{Guilbert-Lepoutre}, A. and others",
  year         = 2024,
  howpublished = "Rapport de prospective INSU-AA, \url{https://prospective-aa.sciencesconf.org/data/rapport_synthese_groupeI1_moyens_inclusion_diversite_egalite.pdf}"
}

@misc{IFOP2024,
  title        = "Baromètre LGBT+ 2024",
  author       = "{Autre cercle} and {IFOP}",
  year         = 2024,
  howpublished = "\url{https://autrecercle.org/barometre-lgbt/}"
}

@misc{MPDF2021,
  title        = "Action 2021--2023",
  author       = "{Mission pour la place des femmes}",
  year         = 2021,
  howpublished = "\url{https://mpdf.cnrs.fr/wp-content/uploads/2024/03/Plaquette-MPDF-8.0.pdf}"
}

@article{Devine2012,
  title     = {Long-term reduction in implicit race bias: A prejudice habit-breaking intervention},
  author    = {Devine, Patricia G. and Forscher, Patrick S. and Austin, Anthony J. and Cox, William T. L.},
  journal   = {Journal of Experimental Social Psychology},
  volume    = {48},
  number    = {6},
  pages     = {1267--1278},
  year      = {2012},
  publisher = {Elsevier},
  doi       = {10.1016/j.jesp.2012.06.003},
  url       = {https://doi.org/10.1016/j.jesp.2012.06.003}
}

@BOOK{Senac2017-he,
  title     = "Les non fr{\`e}res au pays de l'{\'e}galit{\'e}",
  author    = "S{\'e}nac, R{\'e}jane",
  publisher = "Les Presses de Sciences Po",
  year      =  2017,
  language  = "fr"
}

@article{Bian2017,
  title     = {Expectations of brilliance underlie gender distributions across academic disciplines},
  author    = {Bian, Lindsay and Leslie, Sarah-Jane and Cimpian, Andrei},
  journal   = {Science},
  volume    = {355},
  number    = {6323},
  pages     = {389--391},
  year      = {2017},
  publisher = {American Association for the Advancement of Science},
  doi       = {10.1126/science.1261375},
  url       = {https://doi.org/10.1126/science.1261375}
}

@article{Master2021,
	doi = {10.1073/pnas.2100030118},
	url = {https://doi.org/10.1073/pnas.2100030118},
	year = {2021},
	month = {nov},
	publisher = {Proceedings of the National Academy of Sciences},
	volume = {118},
	number = {48},
	author = {Andrei Master and Andrew N. Meltzoff and Sapna Cheryan},
	title = {Gender stereotypes about interests start early and cause gender disparities in computer science and engineering},
	journal = {Proceedings of the National Academy of Sciences}
}

@article{Meltzoff2024,
    author = {Meltzoff, Andrew N. and Gilliam, Walter S.},
    title = {Young Children \&amp; Implicit Racial Biases},
    journal = {Daedalus},
    volume = {153},
    number = {1},
    pages = {65-83},
    year = {2024},
    month = {03},
    issn = {0011-5266},
    doi = {10.1162/daed_a_02049},
    url = {https://doi.org/10.1162/daed\_a\_02049},
    eprint = {https://direct.mit.edu/daed/article-pdf/153/1/65/2345811/daed\_a\_02049.pdf},
}

@article{Chang2023,
title = {Approaches to address bias in letters of recommendation},
journal = {Trends in Pharmacological Sciences},
volume = {44},
number = {6},
pages = {321-323},
year = {2023},
issn = {0165-6147},
doi = {https://doi.org/10.1016/j.tips.2023.03.002},
url = {https://www.sciencedirect.com/science/article/pii/S0165614723000433},
author = {Vivian Y. Chang and Mary Munson and Christina Marie Termini}
}

@article{clance1978imposter,
  title   = {The imposter phenomenon in high achieving women: Dynamics and therapeutic intervention},
  author  = {Clance, Pauline Rose and Imes, Suzanne A.},
  journal = {Psychotherapy: Theory, Research \& Practice},
  volume  = {15},
  number  = {3},
  pages   = {241--247},
  year    = {1978},
  doi     = {10.1037/h0086006},
  url     = {https://doi.org/10.1037/h0086006}
}

@misc{defenseur2021kit,
  author       = {{Défenseur des droits} and Conférence Permanente Égalité Diversité and Ministère de l{'}Enseignement Supérieur et de la Recherche and Association Française des Managers de la Diversité and Jurisup},
  title        = {Kit de prévention des discriminations dans l’enseignement supérieur},
  year         = {2021},
  note         = {Guide publié par le Défenseur des droits et ses partenaires},
  url          = {https://juridique.defenseurdesdroits.fr/index.php?lvl=notice_display&id=42358}
}

@article{ABRAHAM2024348,
title = {Words matter: Experimental evidence from job applications},
journal = {Journal of Economic Behavior \& Organization},
volume = {225},
pages = {348-391},
year = {2024},
issn = {0167-2681},
doi = {https://doi.org/10.1016/j.jebo.2024.06.013},
url = {https://www.sciencedirect.com/science/article/pii/S0167268124002312},
author = {Lisa Abraham and Johannes Hallermeier and Alison Stein}
}

@INPROCEEDINGS{Clavel2024,
       author = {{Clavel}, M. and {Guilbert-Lepoutre}, A. and {Ouazzani}, R. -M. and {Paumard}, T. and {Venot}, O.},
        title = "{Equality Correspondents: the INSU-AA network and its actions}",
    booktitle = {SF2A-2024: Proceedings of the Annual meeting of the French Society of Astronomy and Astrophysics},
         year = 2024,
       editor = {{B{\'e}thermin}, M. and {Bailli{\'e}}, K. and {Lagarde}, N. and {Malzac}, J. and {Ouazzani}, R.~M. and {Richard}, J. and {Venot}, O. and {Siebert}, A.},
        month = dec,
        pages = {17-20},
       adsurl = {https://ui.adsabs.harvard.edu/abs/2024sf2a.conf...17C}
}

@misc{FiveDs,
  author       = {{Right to Be}},
  title        = {Bystander Intervention Training Guide},
  year         = {2017},
  howpublished = {https://righttobe.org/guides/bystander-intervention-training/}
}

\end{document}